\documentclass[preprint, numbers]{deepmark}
\usepackage{amsmath,amssymb,amsfonts,graphicx,booktabs,url}
\usepackage{algorithm, algpseudocode}

\title{Backward Compatibility in Zero-Bit Neural Audio Watermarking}

\author{Admir Krni\'c\dmkcorrauthormark}
\author{Kosta Pavlovi\'c}
\author{Petar Nedi\'c}
\author{\v{Z}eljko Bolevi\'c}
\author{Slavko Kova\v{c}evi\'c}
\author{Igor Djurovi\'c}

\dmkcorrespondence{Admir Krni\'c, \href{mailto:admir.krnic7@gmail.com}{admir.krnic7@gmail.com}}
\affil{DeepMark}

\dmkkeywords{audio watermarking; zero-bit watermarking; backward compatibility; legacy chain; revocation}

\newcommand{\Lpush}{\mathcal{L}_{\text{push}}}
\newcommand{\Ldiv}{\mathcal{L}_{\text{div}}}
\newcommand{\gpush}{\mathbf{g}_{\text{push}}}
\newcommand{\gdiv}{\mathbf{g}_{\text{div}}}

\begin{document}

\maketitle
\thispagestyle{firststyle}

\begin{abstract}
Neural watermarking systems are typically trained as tightly coupled embedder-detector pairs and implicitly assume static deployment. In real-world deployments, embedders and detectors are frequently updated and selectively decoupled for security and versioning purposes, which can break detection in previously embedded content and, more critically, increase false-positive rates on embedders that a given detector is intentionally not intended to recognize.

In this paper the concept of backward compatibility in zero-bit audio watermarking is formalized. A training framework is proposed that pairs a new detector with a sequence of frozen legacy embedders (its legacy chain), using a performance-weighted sampling strategy so that the detector retains accuracy across all earlier versions. This coupling, while necessary for backward compatibility, is shown to induce an embedder-collapse effect that catastrophically degrades the true negative rate (TNR) against out-of-chain embedders. To counteract this, a Angular Diversity Loss (ADL) is introduced that explicitly forces each new embedder to occupy a distinct region of the output space. The combined approach restores near-perfect TNR with no measurable cost to detection accuracy or perceptual audio quality.
\end{abstract}

\section{Introduction}
\label{sec:intro}

Recent advances in generative AI have substantially increased the realism and accessibility of synthetic audio content. State-of-the-art audio generation techniques are capable of producing high-quality speech and audio at scale \cite{valle}. As a result, concerns surrounding large-scale misinformation, fraud and reputational damage have intensified, alongside a broader erosion of trust in digital media.

Audio watermarking, a technique that embeds imperceptible signals at generation time and later recovers them to verify provenance, has emerged as a practical mechanism for attribution and misuse prevention. It is also becoming a legal requirement: the European Union's AI Act (Article~50) mandates that providers of generative AI mark and detect artificially generated content, and its accompanying Code of Practice further requires that a replacement detector be backward compatible, detecting previously marked content with no loss of capability \cite{eu_code_of_practice}.

A line of recent deep-learning approaches jointly trains embedders and detectors, which we survey in Section~\ref{sec:related}. These systems all share an implicit assumption: an embedder and its detector are trained as a single matched pair and deployed together. In practice, watermarking systems must evolve, with new embedder-detector pairs introduced over time to harden against new attacks, improve quality, or reduce cost, while previously distributed content stays in circulation and must remain verifiable. Existing methods provide no explicit mechanism for an updated detector to keep recognizing watermarks from earlier embedders while still rejecting unrelated content, leading to invalidated watermarks or elevated false positives after updates. Re-embedding old content is often infeasible, and maintaining many detector versions at inference is complex and itself raises the false-positive rate.

We call the ordered sequence of legacy embedders $E_1, \ldots, E_{v-1}$, that a newly trained detector $D_v$ must continue to support, the \emph{legacy chain}. A backward-compatible $D_v$ must (i) detect watermarks from all embedders in its chain, (ii) reject unwatermarked audio, and (iii) reject watermarks from embedders outside the chain. Objective (iii) is not merely an accuracy concern; it is what makes a compromised version revocable. If an embedder $E_i$ is leaked, an adversary can stamp arbitrary audio with its watermark. The response is to revoke $E_i$, dropping it from every future detector's chain so that any detector issued after the leak rejects it outright. Without this, a single leak permanently invalidates the watermark, as every newer detector keeps accepting the forgeries. This motivates strict version separation rather than monotone forward acceptance.

In this work, we conduct the first systematic study of backward compatibility in neural audio watermarking. The following contributions are made:
\begin{itemize}
  \item The backward-compatibility objective for zero-bit watermarking is formalized, and the legacy-chain abstraction is introduced together with its security implications.
  \item A training framework is proposed that combines a performance-weighted sampling scheme across frozen legacy embedders with ADL applied to the new embedder.
  \item The framework is evaluated across eight model versions that span architectural changes in both embedders (DenseNet, ViTMAE) and detectors (ResNet, AST), and the ADL is shown to substantially restore TNR against out-of-chain embedders with no measurable degradation in detection accuracy or audio quality.
  \item The same mechanism is shown to enable revocation: a leaked embedder can be dropped from the legacy chain and reliably rejected by detectors trained without it.
\end{itemize}

\section{Related Work}
\label{sec:related}

Deep-learning-based audio watermarking jointly trains an embedder and a detector to balance imperceptibility and robustness. RobustDNN \cite{robustdnn} introduced the basic recipe, WavMark \cite{wavmark}, AudioSeal \cite{audioseal}, Timbre Watermarking \cite{timbre}, and SilentCipher \cite{silentcipher} subsequently improved robustness to common signal-processing attacks. VoiceMark \cite{voicemark} extends robustness to zero-shot voice-cloning attacks. All of these systems train and deploy embedders and detectors as fixed pairs and provide no explicit mechanism for cross-version compatibility.

Petrov \emph{et al.} \cite{petrov_coexistence} study the coexistence and ensembling of independently trained watermarks. Their setting differs from ours in two important ways: embedder-detector pairs are trained independently (with no shared parameters or training signal), and there is no obligation for a later detector to recognize watermarks from an earlier one. As a consequence, their reported interference is minimal. We instead consider the coupled-training setting required for backward compatibility, and show that this coupling introduces failure modes that independent training does not exhibit.

Continual learning addresses sequential task acquisition without catastrophic forgetting \cite{parisi_continual}, through methods such as Elastic Weight Consolidation \cite{ewc}, Learning without Forgetting \cite{lwf}, and Synaptic Intelligence \cite{synaptic}. These approaches assume a single, architecturally fixed model updated over time and therefore do not accommodate the architectural changes that accompany real-world watermarking updates.

\section{Problem Formulation}
\label{sec:problem}

We first review zero-bit watermarking and then formalize the backward-compatibility objective on top of it.

\textbf{Zero-Bit Watermarking.} Zero-bit watermarking aims to decide whether a digital signal contains a watermark. Given an input signal $x \in \mathcal{X}$, an embedder $E$ produces a watermarked signal $x_w = E(x)$, where the perturbation $E(x) - x$ is imperceptible. A detector $D$ then produces a real-valued detection score $D(x_w) \in \mathbb{R}$, with larger positive values indicating the presence of a watermark and smaller (negative) values indicating its absence. A binary decision is obtained by comparing the score against a threshold. The pair $(E, D)$ is typically trained end-to-end: the detector learns to distinguish watermarked from unwatermarked audio, while the embedder learns to introduce minimal perturbations that the detector can reliably detect.

\textbf{Backward-Compatibility Objective.} We associate every training run with a discrete version index $v$. For the initial version, $v = 1$, a single pair $(E_1, D_1)$ is trained end-to-end. For each subsequent version $v > 1$, a new embedder $E_v$ is trained jointly with a new detector $D_v$. The legacy chain at version $v$ is the set of earlier embedders $\{E_i : i < v\}$, minus a revocation set $R$ of embedders deliberately dropped (e.g.\ a leaked one). The legacy chain is kept frozen during training, and $D_v$ is trained against it together with $E_v$, so its compatible (in-chain) set is $\{E_i : i \le v\} \setminus R$. Every other embedder is out-of-chain: future versions $E_j$ with $j > v$ that did not yet exist, and any revoked $E_i \in R$.

Formally, the detector $D_v$ at version $v$ must satisfy:
\begin{itemize}
  \item $D_v$ correctly detects watermarks produced by every embedder from set: $\{E_i : i \le v\} \setminus R$ (high TPR within the chain).
  \item $D_v$ rejects raw audio (high TNR on clean inputs).
  \item $D_v$ rejects watermarks produced by any embedder outside its legacy chain (high TNR against out-of-chain embedders): both future versions $E_j$ with $j > v$ that did not yet exist at training, and any embedder from $R$.
\end{itemize}

The last objective we refer to as the \emph{version boundary}: the line a detector holds between the in-chain embedders it accepts and the out-of-chain ones it rejects, whether future or revoked. This is what distinguishes backward compatibility from standard multi-embedder training: an out-of-chain embedder is not merely absent, it must be actively rejected, which is precisely what makes revoking a leaked embedder possible.

\section{Method}
\label{sec:method}

Although backward compatibility is a general concern, we focus exclusively on zero-bit audio watermarking in this work. Our method has three components: a backward-compatible training setup, a performance-weighted sampling strategy that allocates training iterations across the legacy chain, and a diversity loss that prevents the new embedder from drifting toward its predecessors.

\textbf{Backward-Compatible Training Setup.} Embedders operate in the short-time Fourier transform (STFT) domain. The watermark is added to the magnitude spectrogram within the $[500, 4000]$~Hz embedding band and the time-domain signal is reconstructed via inverse STFT using the original phase. Detectors mirror the embedder, they compute the STFT and a mel spectrogram restricted to the same embedding band, then feed it to the detection network.

We extend the idea from SilentCipher \cite{silentcipher} of enforcing a hard constraint on the maximum perceptual distortion, which lets us train with a detection-only loss. This distortion constraint drives all embedders to a consistent quality level (PESQ 4.3--4.5). Because every model then sits at essentially the same perceptual quality, we can compare models by detection accuracy alone.

For the detection objective we adopt the push loss introduced by AWARE \cite{aware}, which augments the standard detection MSE with a term that widens the margin between watermarked and unwatermarked frames:
\begin{equation}
\Lpush = \mathrm{MSE}(\hat{y}, y) - \lambda \left( \frac{1}{|y \ge 0|} \sum_{y \ge 0} \hat{y} + \frac{1}{|y < 0|} \sum_{y < 0} (-\hat{y}) \right),
\label{eq:push}
\end{equation}
where the second term pushes positive logits (watermarked frames) up and negative logits (unwatermarked frames) down, scaled by the push weight $\lambda$.

For version $v = 1$, training follows the standard procedure: a single embedder-detector pair is trained end-to-end with the push loss alone. For versions $v > 1$, two things change with respect to the $v = 1$ recipe. First, the new embedder $E_v$ and detector $D_v$ may use a different architecture from any previous version. This is by design, since real watermarking deployments evolve their architectures over time (Section~\ref{sec:setup}). Second, the training objective is extended with the diversity loss introduced below, which is needed to keep the new embedder distinguishable from the legacy chain and thereby preserve the version boundary.

At each iteration, the detector $D_v$ is trained on the output of a single embedder sampled from the full chain $\{E_1, \ldots, E_v\} \setminus R$, chosen by the performance-weighted sampling scheme described in the next paragraph. Because the legacy embedders are frozen and each embedder is drawn independently by this scheme, the training of $D_v$ does not depend on the order of the embedders within the chain. The version index reflects only the order in which the embedders were created.

\textbf{Performance-Weighted Sampling.} Let's denote with $\mathcal{C}$ the chain of embedders the detector is trained to detect $\{E_1, \ldots, E_v\} \setminus R$. At each training iteration we sample a single embedder $E_i \in \mathcal{C}$ according to a probability distribution $\mathbf{p}$ over $\mathcal{C}$, and perform the forward and backward passes for the pair $(E_i, D_v)$. The distribution $\mathbf{p}$ is initialized to uniform. After each validation step we obtain a per-embedder accuracy $A_i \in [0, 1]$ for every $E_i \in \mathcal{C}$ and update $\mathbf{p}$ via a temperature-controlled softmax \cite{hinton_distill} over negated accuracies:
\begin{equation}
p_i = \frac{\exp(-k \cdot A_i)}{\sum_{E_j \in \mathcal{C}} \exp(-k \cdot A_j)},
\label{eq:sampling}
\end{equation}
where $k > 0$ controls how aggressively the budget is shifted toward weaker embedders. The scheme adaptively emphasizes the embedders on which the detector currently performs worst. Based on our experiments $k = 8$ yielded the best results.

\textbf{Embedder Collapse and the Need for a Diversity Loss.} Performance-weighted sampling together with push loss is sufficient to obtain high TPR within the legacy chain, but it leaves the new embedder $E_v$ entirely free to position its residual watermark anywhere in the signal space. The detector $D_v$ is itself trained from scratch, what is fixed are the legacy embedders $E_1, \ldots, E_{v-1}$, whose residuals are frozen throughout training while $E_v$'s residual is the only one free to move. The easiest path to a low detection loss is therefore for $E_v$ to drift toward the frozen legacy embedders' residual distributions, so that $D_v$ can satisfy every in-chain pair with a single decision boundary. We refer to this failure mode as \emph{embedder collapse}: the new embedder's residual ends up implicitly close to those of the frozen legacy embedders, destroying the version boundary.

\textbf{Angular Diversity Loss.} We treat the mel spectrogram of the audio watermarked by each embedder as a ``class'' on the unit sphere and separate these classes with an additive angular margin, inspired by the margin formulation from ArcFace \cite{arcface}. The setting, however, is not the fixed-identity classification ArcFace was designed for: an embedder has no identity vector to learn, and only $E_v$ is trainable while the legacy embedders are frozen. The ADL adapts the angular margin to this setting with two changes.

First, we remove ArcFace's learnable per-class prototypes. Let $f_i(x)$ denote the batch mean of the mel feature of the audio watermarked by embedder $E_i$. Each class center is then the normalized batch mean $\mu_i = \bar{f}_i / \lVert \bar{f}_i \rVert$, recomputed on every forward pass. Second, since only $E_v$ moves, the margin is applied only to the mel features of the new embedder's output. We denote by $\cos\theta_i^{(b)} = \langle f_v(x_b), \mu_i \rangle$ the cosine similarity between mel features of the sample $b$ produced by the new embedder and class center $i$. Let $\mathcal{N}_v$ be the set of class centers from which $E_v$ is repelled, typically $\mathcal{N}_v$ contains all predecessors $\{E_i : i < v\}$, with the option of removing revoked embedders, as in our revocation experiments. The loss is the additive-margin cross-entropy that pulls every sample toward its own class $v$:
\begin{equation}
\Ldiv = -\frac{1}{B} \sum_{b} \log \frac{e^{s \cos(\theta_v^{(b)} + m)}}{e^{s \cos(\theta_v^{(b)} + m)} + \sum_{i \in \mathcal{N}_v} e^{s \cos\theta_i^{(b)}}}.
\label{eq:adl}
\end{equation}
The margin $m$ pushes each sample of $E_v$ at least $m$ radians away from every class center in $\mathcal{N}_v$, while the scale $s$ controls the softmax sharpness.

The proposed loss is essential for preserving the separation between model versions. Without it, the embedding spaces of the legacy and new embedders collapses, eliminating the intended version boundary. The mel spectrogram is a robust representation in which perceptually similar signals produce very similar mel features, so a strong margin or sharp scale would destabilize training. This is why weak parameters ($m=0.1$, $s=8$) suffice: the loss still enforces the separation, but needs only a gentle push.

\textbf{Gradient Normalization.} Combining $\Lpush$ and $\Ldiv$ through a fixed scalar weighting is brittle: the two losses operate on very different scales so one term tends to dominate the gradient. We instead balance them with a per-step gradient-norm scaling inspired by GradNorm \cite{gradnorm}.

At each step we compute the per-loss gradients with respect to the new embedder's parameters, $\gpush = \nabla \Lpush$ and $\gdiv = \nabla \Ldiv$, and their $L_2$ norms. We then rescale both to a common target norm $\bar{n} = \tfrac{1}{2}\left(\lVert \gpush \rVert + \lVert \gdiv \rVert\right)$, and sum the result:
\begin{equation}
\mathbf{g} = \frac{\bar{n}}{\lVert \gpush \rVert + \epsilon}\, \gpush + \frac{\bar{n}}{\lVert \gdiv \rVert + \epsilon}\, \gdiv.
\label{eq:gradnorm}
\end{equation}
Dividing each gradient by its own norm removes raw magnitude so both losses count equally, while the shared factor $\bar{n}$ rescales them to a common, representative size instead of unit norm.

\section{Experimental Setup}
\label{sec:setup}

This section describes the data, model architectures and evaluation protocol.

\textbf{Dataset.} Training data is drawn from the Mozilla Common Voice corpus \cite{commonvoice}, with all training audio resampled to 44.1~kHz during preprocessing. The test set consists of audio files sampled from three corpora, LibriSpeech \cite{librispeech}, LJSpeech \cite{ljspeech}, and Mozilla Common Voice, and spans a range of native sampling rates, reflecting a realistic and heterogeneous evaluation condition.

\textbf{Model Architectures.} We evaluate two model families with complementary inductive biases: a convolutional family (DenseNet \cite{densenet} embedder with a ResNet \cite{resnet} detector) and a Transformer family (ViTMAE \cite{vit, mae} embedder with an AST \cite{ast} detector). Each architecture uses a fixed template, across versions we vary only its activation function and its depth, holding every other hyperparameter (widths, kernel sizes, attention heads, feature dimensions) constant. The depth knob is specific to each architecture: dense blocks for DenseNet, encoder/decoder layers for ViTMAE, residual stacks for ResNet, and encoder layers for AST. Table~\ref{tab:configs} gives the exact per-version configuration.

\begin{table}[!htbp]
\centering
\small
\caption{Model configurations per version.}
\label{tab:configs}
\begin{tabular}{lll}
\toprule
\textbf{Version} & \textbf{Embedder} & \textbf{Detector} \\
\midrule
$v_1$ & DenseNet: 1 block, ReLU   & ResNet: 1 stack, ReLU \\
$v_2$ & DenseNet: 1 block, Swish  & ResNet: 1 stack, Swish \\
$v_3$ & DenseNet: 1 block, GELU   & ResNet: 1 stack, GELU \\
$v_4$ & DenseNet: 2 blocks, GELU  & ResNet: 2 stacks, GELU \\
$v_5$ & ViTMAE: 1 enc, 2 dec, ReLU  & AST: 4 layers, ReLU \\
$v_6$ & ViTMAE: 1 enc, 2 dec, Swish & AST: 4 layers, Swish \\
$v_7$ & ViTMAE: 1 enc, 2 dec, GELU  & AST: 4 layers, GELU \\
$v_8$ & ViTMAE: 2 enc, 3 dec, GELU  & AST: 6 layers, GELU \\
\bottomrule
\end{tabular}
\end{table}

The version-invariant hyperparameters are as follows. \emph{DenseNet}: $64$ initial filters, dense blocks of $16$ bottleneck layers ($3{\times}3$ convolutions, growth rate $64$, bottleneck reduction $4$). \emph{ViTMAE}: token dimension $128$, one attention head, feed-forward ratio $4$, dropout $0.1$. \emph{ResNet}: $64$ initial filters (kernel size $7$), stacks of four $3{\times}3$ residual blocks, a three-layer LSTM head ($128/256/128$), and a dilated output convolution ($256$ channels). \emph{AST}: hidden size $128$, four attention heads, $16{\times}16$ patches at stride $10$, dropout $0.1$.

\textbf{Modeling realistic architectural drift.} We train a sequence of eight versions ($v_1$ to $v_8$), one model, i.e.\ one embedder-detector pair, per version. Over time, practitioners swap in newer activation functions, enlarge networks, and eventually replace entire architecture families with newer ones. Our version sequence is designed to mimic this trajectory. Concretely, we choose architectures along two orthogonal axes:
\begin{enumerate}
  \item \textbf{Activation rotation.} Within a fixed embedder/detector size, we rotate the activation function in the order in which it became standard in the deep-learning literature: ReLU \cite{relu} first, then Swish \cite{swish}, then GELU \cite{gelu}. This mimics a team that periodically swaps the activation function for a more recent one without otherwise changing the network.
  \item \textbf{Capacity step-ups.} After the activation rotation has cycled through ReLU/Swish/GELU at the original size, we step up the network's capacity (more dense blocks / more encoder-decoder layers) while keeping the most recent activation. This mimics a team that, after exhausting easy activation upgrades, decides to scale the model.
\end{enumerate}

The same two-axis schedule is applied first to the convolutional family ($v_1$--$v_4$) and then repeated on the Transformer family ($v_5$--$v_8$), since Transformer-based audio models entered the field later than convolutional ones. The transition from $v_4$ to $v_5$ simulates the discrete jump that occurs when a deployment switches to a fundamentally newer architecture family rather than enlarging the existing one.

\textbf{Training Details.} All models are trained with Adam at a learning rate of $3 \times 10^{-4}$, decayed by an exponential schedule ($\gamma = 0.95$), with a batch size of $128$ and gradient-norm clipping at $0.1$.

\textbf{Evaluation Metrics.} For each detector $D_v$ we measure: TPR on in-chain pairs, TNR on unwatermarked audio and on out-of-chain pairs, respectively.

Results are presented as accuracy matrices, evaluated on the test set described above. A binary decision requires a confidence threshold $\tau$. We report two experiments: a \emph{base setup} with performance-weighted sampling and the detection loss, and the \emph{full method} that adds the ADL on top. Both use all eight versions with no revocation ($R = \emptyset$). Revocation is evaluated separately.

\section{Results}
\label{sec:results}

We fix the confidence threshold at $\tau = 0.99$, which gives the best TPR and TNR for our objectives across both experiments.

\textbf{Unwatermarked Audio.} We evaluate every detector on the unwatermarked test set and measure the TNR. All detectors, across both training approaches, achieve $100\%$ TNR.

\begin{figure}[!htbp]
  \centering
  \includegraphics[width=0.6\linewidth]{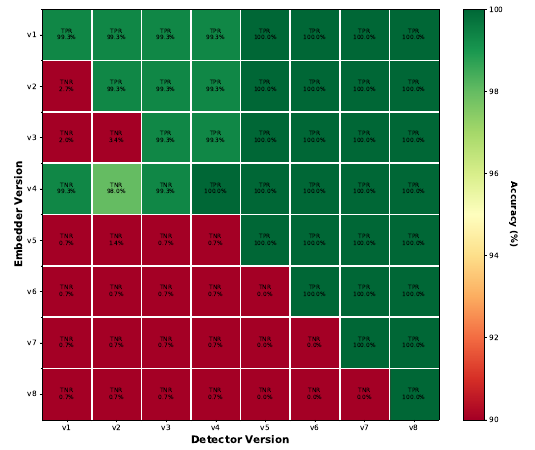}
  \caption{Accuracy matrix for the base setup (no diversity loss). Rows index embedder versions, columns index detector versions. The main diagonal and entries above it show TPR entries below show TNR. In-chain TPR is near-perfect, but the lower triangle collapses: the detector accepts almost all out-of-chain watermarks as its own.}
  \label{fig:acc_base}
\end{figure}

\textbf{Results Without Diversity Loss.} Our base setup pairs each new detector with the legacy chain using performance-weighted sampling and the detection loss alone, with no diversity term. Fig.~\ref{fig:acc_base} shows its accuracy matrix.

In-chain TPR is near-perfect, between $99.3\%$ and $100\%$ (mean $99.8\%$), so performance-weighted sampling alone lets each detector serve its whole chain. The version boundary, however, fails catastrophically: mean out-of-chain TNR is only $11.3\%$, with $24$ of $28$ incompatible pairs at $\le 3\%$. The collapse is total within the ViTMAE block ($0\%$ TNR) and near-total across cross-family transitions ($0.7$--$1.4\%$). The sole exception is embedder $v_4$, rejected by earlier detectors at high TNR.

\textbf{Residual-Similarity Evidence for Embedder Collapse.} The accuracy matrix shows that the version boundary collapses without a diversity loss, but not \emph{why}. Our hypothesis is that, with no pressure to differ, a new embedder learns watermark residuals that resemble its predecessors', so their detectors accept it as one of their own.

The \emph{residual} of an embedder is the perturbation it adds to the input, $E(x) - x$, i.e.\ the watermark signal itself. To test the hypothesis we take each incompatible pair $(E_M, D_N)$ with $M > N$ (the lower triangle of Fig.~\ref{fig:acc_base}) and compare two quantities: its TNR and
\begin{equation}
\bar{c}_{M,N} = \frac{1}{N} \sum_{k=1}^{N} c(M, k),
\label{eq:cbar}
\end{equation}
where $c(M, k)$ is the averaged cosine similarity between the residuals of $E_M$ and $E_k$. A high $\bar{c}_{M,N}$ means $E_M$'s watermark looks, in residual space, like the watermarks $D_N$ already accepts. If the hypothesis holds, TNR should fall as $\bar{c}_{M,N}$ rises.

\begin{table}[!htbp]
\centering
\small
\caption{Out-of-chain TNR grouped by mean residual similarity $\bar{c}_{M,N}$.}
\label{tab:tnr_residual}
\begin{tabular}{lcc}
\toprule
$\bar{c}_{M,N}$ & $n$ & TNR (\%) \\
\midrule
$[0.45, 0.55)$ & 3  & \textbf{98.9} \\
$[0.55, 0.65)$ & 1  & 2.7 \\
$[0.65, 0.75)$ & 15 & 0.9 \\
$[0.75, 0.85)$ & 9  & 0.4 \\
\bottomrule
\end{tabular}
\end{table}

Across all pairs, $\bar{c}_{M,N}$ and TNR are strongly negatively correlated (Pearson $r = -0.86$, Spearman $\rho = -0.73$ both $p < 10^{-4}$). The more a new embedder's residual resembles those in the detector's chain, the less reliably that detector rejects it. Binning the pairs by $\bar{c}_{M,N}$ (Table~\ref{tab:tnr_residual}) makes the effect concrete: TNR stays near-perfect only in the lowest-similarity bin and falls off a cliff the moment $\bar{c}_{M,N}$ crosses ${\sim}0.55$. This confirms that residual closeness directly drives the failed rejections and motivates a loss that keeps successive embedders apart in residual space.

\begin{figure}[!htbp]
  \centering
  \includegraphics[width=0.6\linewidth]{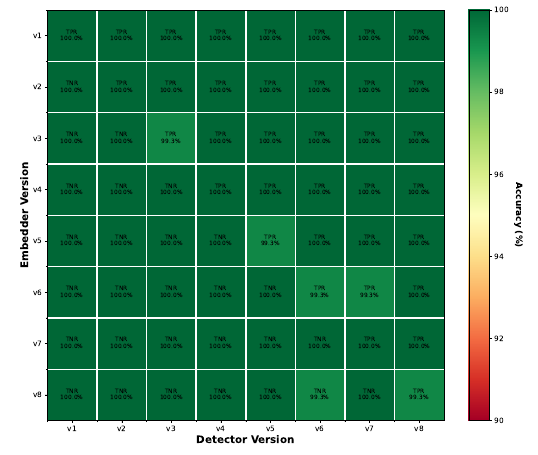}
  \caption{Accuracy matrix for the full method (with the ADL). Same layout as Fig.~\ref{fig:acc_base}. In-chain TPR is unchanged, while the lower triangle recovers to near-perfect TNR.}
  \label{fig:acc_adl}
\end{figure}

\textbf{Results With Diversity Loss.} The full method adds the ADL on top of the base setup, keeping performance-weighted sampling and the detection loss and combining losses through gradient normalization. Fig.~\ref{fig:acc_adl} presents the resulting accuracy matrix.

Compatibility is preserved: in-chain TPR stays between $99.3$ and $100\%$ (mean $99.9\%$), matching the base setup. The diversity loss is therefore free in compatibility terms.

Out-of-chain rejection, the failure mode of the base setup, is recovered: mean TNR rises from $11.3\%$ to $99.98\%$ ($27$ of $28$ incompatible pairs at $100\%$, the last at $99.3\%$), with no residual failure cluster. The version boundary holds within the convolutional family ($v_1$--$v_4$), within the transformer family ($v_5$--$v_8$), and across the $v_4 \rightarrow v_5$ family switch, enforcing separation across both within- and cross-family transitions.

\begin{table}[!htbp]
\centering
\small
\caption{Revocation: TNR (\%) against the revoked embedder.}
\label{tab:revocation}
\begin{tabular}{lccc}
\toprule
Revoked & No div. & Div., $E_i$ unused & Div., $E_i$ pushed \\
\midrule
$v_3$ (conv.)   & 0.7 & 24.3 & \textbf{100.0} \\
$v_5$ (transf.) & 0.0 & 95.9 & \textbf{100.0} \\
\bottomrule
\end{tabular}
\end{table}

\textbf{Revocation.} Revoking a leaked embedder $E_i$ means dropping it from every future detector's legacy chain, which is considered successful if a detector trained \emph{without} $E_i$ still rejects it. Starting from the chain of versions $v_1$ to $v_7$, we revoke a single embedder, either a convolutional-family one ($v_3$) or a transformer-family one ($v_5$), and for each retrain a new embedder--detector pair on the six surviving versions ($v_1$ to $v_7$ excluding the revoked one) under three conditions: (i) the baseline with no diversity loss, (ii) the diversity loss with $E_i$ unused and (iii) the diversity loss with $E_i$ added to the diversity term only (forcing the new embedder to differ from it) while kept out of the detection loss, so the detector is never trained to accept it. Table~\ref{tab:revocation} reports TNR against $E_i$. The baseline barely rejects the revoked embedder, and merely omitting it from a diversity-trained chain is unreliable. Only when $E_i$ is pushed away by the diversity loss does the new embedder produce distinct residuals, and the detector rejects the revoked embedder fully in both families.

\textbf{Third-Party Watermarks.} In production, our detectors must not accept third-party watermarks, and our watermarked content must not be accepted by third-party detectors. We tested both directions against the publicly released zero-bit AudioSeal model \cite{audioseal}, using all eight pair of models trained with the full method: (i) the audio watermarked by each of our embedders is passed to the AudioSeal detector, and (ii) the audio watermarked by the AudioSeal embedder is passed to each of our detectors. Because AudioSeal operates at 16~kHz, audio is resampled to 16~kHz at its input. 

We use AudioSeal's default decision threshold of $0.5$. For every one of the eight versions, and in both directions, not a single file is detected, so TNR is $100\%$ for every pair. This agrees with the negligible interference that Petrov \emph{et al.}~\cite{petrov_coexistence} report for independently trained watermarks.

\section{Robustness Analysis}
\label{sec:robustness}

Maximizing robustness to signal-processing and codec attacks is out of scope, and none of our models are trained with attack augmentation. We ask a narrower question: does backward compatibility itself survive distortion? A deployed watermark is transmitted and edited before detection, so a framework that held only on clean audio would be of little practical use. We therefore test, rather than optimize for, robustness, re-running the evaluation under the DeepMark benchmark \cite{deepmark_benchmark}. Table~\ref{tab:attacks} reports the min/max/mean TPR and TNR per attack for both training conditions; the version boundary is preserved under attack.

\begin{table}[!htbp]
\centering
\footnotesize
\setlength{\tabcolsep}{3.5pt}
\caption{Per-attack TPR (\%) and TNR (\%) at $\tau = 0.99$.}
\label{tab:attacks}
\begin{tabular}{l*{12}{c}}
\toprule
 & \multicolumn{6}{c}{\textbf{Baseline}} & \multicolumn{6}{c}{\textbf{Ours}} \\
\cmidrule(lr){2-7} \cmidrule(lr){8-13}
 & \multicolumn{3}{c}{TPR (\%)} & \multicolumn{3}{c}{TNR (\%)} & \multicolumn{3}{c}{TPR (\%)} & \multicolumn{3}{c}{TNR (\%)} \\
\cmidrule(lr){2-4} \cmidrule(lr){5-7} \cmidrule(lr){8-10} \cmidrule(lr){11-13}
\textbf{Attack} & min & max & avg & min & max & avg & min & max & avg & min & max & avg \\
\midrule
Additive Noise     & 25.0 & 69.5  & \textbf{52.7} & 26.3 & 100.0 & 49.5 & 16.8 & 76.3  & 44.3          & 100.0 & 100.0 & \textbf{100.0} \\
Bandstop           & 66.2 & 66.2  & \textbf{66.2} & 33.7 & 100.0 & 41.0 & 65.5 & 66.2  & 66.1          & 99.3  & 100.0 & \textbf{99.9} \\
Crop Beginning     & 99.3 & 100.0 & 99.7          & 0.0  & 100.0 & 11.6 & 98.6 & 100.0 & 99.7          & 99.3  & 100.0 & \textbf{99.9} \\
Crop Random        & 98.6 & 100.0 & 99.7          & 0.0  & 100.0 & 11.7 & 98.6 & 100.0 & \textbf{99.8} & 100.0 & 100.0 & \textbf{100.0} \\
Cut Samples        & 99.3 & 100.0 & \textbf{99.7} & 0.0  & 99.3  & 11.6 & 97.9 & 100.0 & 99.6          & 100.0 & 100.0 & \textbf{100.0} \\
Echo               & 98.6 & 100.0 & \textbf{99.7} & 0.0  & 100.0 & 11.5 & 96.6 & 100.0 & 99.4          & 98.6  & 100.0 & \textbf{99.9} \\
Flanger            & 96.6 & 100.0 & \textbf{99.3} & 0.0  & 100.0 & 13.0 & 92.5 & 100.0 & 98.7          & 100.0 & 100.0 & \textbf{100.0} \\
Flip Samples       & 96.6 & 100.0 & \textbf{99.4} & 0.0  & 100.0 & 12.1 & 95.9 & 100.0 & 99.0          & 99.3  & 100.0 & \textbf{99.9} \\
Gaussian Noise     & 75.0 & 98.6  & \textbf{87.9} & 1.3  & 99.3  & 21.4 & 58.7 & 97.2  & 82.9          & 99.3  & 100.0 & \textbf{99.9} \\
Highpass           & 99.3 & 100.0 & 99.7          & 0.0  & 99.3  & 11.1 & 98.6 & 100.0 & \textbf{99.8} & 99.3  & 100.0 & \textbf{99.9} \\
Inverted Time-Str. & 98.6 & 100.0 & \textbf{99.8} & 0.0  & 100.0 & 11.1 & 97.2 & 100.0 & 99.7          & 72.9  & 100.0 & \textbf{98.2} \\
LPC                & 98.6 & 100.0 & 99.7          & 0.0  & 99.3  & 11.2 & 98.6 & 100.0 & \textbf{99.8} & 100.0 & 100.0 & \textbf{100.0} \\
Lowpass            & 99.3 & 100.0 & 99.8          & 0.0  & 99.3  & 11.2 & 99.3 & 100.0 & 99.8          & 100.0 & 100.0 & \textbf{100.0} \\
MP3 Compression    & 97.2 & 100.0 & \textbf{99.3} & 0.0  & 99.3  & 11.6 & 97.2 & 100.0 & 99.0          & 99.3  & 100.0 & \textbf{99.9} \\
Network Transm.    & 62.8 & 100.0 & \textbf{88.6} & 0.0  & 100.0 & 21.4 & 23.6 & 95.9  & 74.4          & 99.3  & 100.0 & \textbf{99.9} \\
PCM Quantization   & 98.6 & 100.0 & \textbf{99.5} & 0.0  & 99.3  & 11.3 & 97.9 & 100.0 & 99.4          & 99.3  & 100.0 & \textbf{99.9} \\
Pink Noise         & 26.3 & 78.3  & \textbf{57.2} & 20.2 & 100.0 & 46.2 & 18.2 & 85.8  & 50.1          & 100.0 & 100.0 & \textbf{100.0} \\
Pitch Shift        & 98.6 & 100.0 & 99.6          & 0.0  & 100.0 & 11.5 & 97.9 & 100.0 & 99.6          & 93.2  & 100.0 & \textbf{99.6} \\
Quantization       & 73.6 & 99.3  & \textbf{89.3} & 0.6  & 100.0 & 20.2 & 56.7 & 96.6  & 81.9          & 97.2  & 100.0 & \textbf{99.8} \\
Replacement        & 97.2 & 100.0 & \textbf{99.0} & 0.6  & 99.3  & 12.4 & 93.9 & 100.0 & 98.3          & 100.0 & 100.0 & \textbf{100.0} \\
Resampling         & 99.3 & 100.0 & 99.7          & 0.0  & 99.3  & 11.2 & 98.6 & 100.0 & \textbf{99.8} & 100.0 & 100.0 & \textbf{100.0} \\
STFT Quantization  & 70.2 & 97.9  & \textbf{93.2} & 1.3  & 99.3  & 13.3 & 42.5 & 100.0 & 81.7          & 91.2  & 100.0 & \textbf{98.8} \\
Sign Inversion     & 99.3 & 100.0 & 99.8          & 0.0  & 99.3  & 11.2 & 99.3 & 100.0 & 99.8          & 99.3  & 100.0 & \textbf{99.9} \\
Smoothing          & 96.6 & 100.0 & \textbf{98.9} & 0.0  & 100.0 & 13.8 & 66.2 & 100.0 & 93.5          & 85.1  & 100.0 & \textbf{98.2} \\
Time Stretch       & 98.6 & 100.0 & \textbf{99.8} & 0.0  & 100.0 & 11.3 & 97.9 & 100.0 & 99.7          & 95.2  & 100.0 & \textbf{99.6} \\
Collusion          & 94.5 & 100.0 & \textbf{98.2} & 0.0  & 100.0 & 13.9 & 92.5 & 100.0 & 97.6          & 99.3  & 100.0 & \textbf{99.9} \\
Zero-Cross Inserts & 99.3 & 100.0 & \textbf{99.8} & 0.0  & 100.0 & 11.4 & 98.6 & 100.0 & 99.7          & 99.3  & 100.0 & \textbf{99.9} \\
\bottomrule
\end{tabular}
\end{table}

\textbf{The version boundary survives attack.} The baseline's TNR was already broken on clean audio and stays broken under every attack (mean 11--14\%), while ours was near-perfect on clean audio and stays that way, with mean TNR $\ge 98\%$ on all $27$ attacks and exactly $100\%$ on $21$.

\textbf{TPR is essentially unchanged.} In-chain detection tracks the baseline across attacks, confirming that adding the diversity loss has a negligible cost in compatibility under distortion.

\section{Conclusion}
\label{sec:conclusion}

We presented the first systematic study of backward compatibility in zero-bit deep-learning audio watermarking. We formalized the backward-compatibility objective via the legacy-chain abstraction, identified the embedder-collapse failure mode that breaks version boundaries when a new detector is trained against frozen legacy embedders, and proposed a training framework that combines performance-weighted sampling and push loss together with ADL balanced by gradient normalization. Across eight versions and two architecture families, the method restores out-of-chain TNR without degrading within-chain detection or perceptual quality, and the restored version boundary holds under distortion. The same mechanism enables revocation: a leaked embedder dropped from the chain is reliably rejected by detectors trained without it. We see the legacy chain as a useful abstraction for future watermarking systems that must evolve under deployment constraints, and we leave robustness under adversarial attack and multibit extensions to future work.

\bibliographystyle{unsrtnat}
\bibliography{refs}

\end{document}